\documentclass[useAMS,usenatbib]{mn2e} 
\usepackage{aas_macros}
\usepackage{graphics}
\usepackage[pdftex]{graphicx}
\usepackage{epstopdf}
\usepackage{epsfig}  
\usepackage{natbib} 
\usepackage{float}
\usepackage{amsmath}
\usepackage{times}
\usepackage[varg]{txfonts}
\usepackage{verbatim} 
\newcommand{\bl}{\boldsymbol}
\newcommand{\tm}{\textrm}

\newcommand{\hMpc}{{\ifmmode{h^{-1}{\rm Mpc}}\else{$h^{-1}$Mpc}\fi}}
\newcommand{\hkpc}{{\ifmmode{h^{-1}{\rm kpc}}\else{$h^{-1}$kpc}\fi}}
\newcommand{\hMsun}{{\ifmmode{h^{-1}{\rm {M_{\odot}}}}\else{$h^{-1}{\rm{M_{\odot}}}$}\fi}}

\def\lesssim{\mathrel{\hbox{\rlap{\hbox{\lower4pt\hbox{$\sim$}}}\hbox{$<$}}}}
\def\gtrsim{\mathrel{\hbox{\rlap{\hbox{\lower4pt\hbox{$\sim$}}}\hbox{$>$}}}}

\title[Reverse Zeldovich Approximation]
      {Reconstructing cosmological initial conditions from galaxy peculiar velocities. I. Reverse Zeldovich Approximation}
\author[Doumler et al.] 
{Timur Doumler$^{1,2}$, Yehuda Hoffman$^{3}$, H\'el\`ene Courtois$^{1}$\thanks{E-mail: h.courtois@ipnl.in2p3.fr}, and Stefan Gottl\"ober$^{2}$    \\
  $^1$Universit\'e Lyon 1, CNRS/IN2P3, Institut de Physique Nucl\'eaire de Lyon, 69622 Villeurbanne, France\\
  $^2$Leibniz-Institut f\"ur Astrophysik Potsdam, An der Sternwarte 16, 14482 Potsdam, Germany\\
  $^3$Racah Institute of Physics, Hebrew University, Jerusalem 91904, Israel\\
  }

\begin{document}

\date{September , 2012}

\pagerange{\pageref{firstpage}--\pageref{lastpage}} \pubyear{2012}

\maketitle

\label{firstpage}


\begin{abstract}
We propose a new method to recover the cosmological initial conditions of the presently observed galaxy distribution, which can serve to run constrained simulations of the Local Universe. Our method, the Reverse Zeldovich Approximation (RZA), can be applied to radial galaxy peculiar velocity data and extends the previously used Constrained Realizations (CR) method by adding a Lagrangian reconstruction step. The RZA method consists of applying the Zeldovich approximation in reverse to galaxy peculiar velocities to estimate the cosmic displacement field and the initial linear matter distribution from which the present-day Local Universe evolved. We test our method with a mock survey taken from a cosmological simulation. We show that the halo peculiar velocities at $z=0$ are close to the linear prediction of the Zeldovich approximation, if a grouping is applied to the data to remove virial motions. We find that the addition of RZA to the CR method significantly improves the reconstruction of the initial conditions. The RZA is able to recover the correct initial positions of the velocity tracers with a median error of only 1.36 Mpc/$h$ in our test simulation. For realistic sparse and noisy data, this median increases to 5 Mpc/$h$. This is a significant improvement over the previous approach of neglecting the displacement field, which introduces errors on a scale of 10 Mpc/$h$ or even higher. Applying the RZA method to the upcoming high-quality observational peculiar velocity catalogues will generate much more precise constrained simulations of the Local Universe.
\end{abstract}


\section{Introduction}
\label{sec:introduction}

\subsection{Constrained simulations}

In the concordance $\Lambda$CDM model of structure formation, the cosmic web of our Universe evolved through gravitational interactions from initial conditions (ICs) which constitute a linear Gaussian random field of density perturbations \citep{Kolb1990}. These initial conditions completely define the assembly history of the contemporary large-scale structure (LSS) of the Universe, and with the knowledge of the theory of gravity, the galaxy formation process, and the basic cosmological parameters, the formation and evolution of the LSS can be modelled. Today, the most successful method to do so are numerical $N$-body simulations (e.g.\ \citealt{Springel2005b,Springel2008,Teyssier2009,Klypin2011}). These simulations reached an impressive dynamical range in mass and length scales and can be applied to study many different aspects of cosmological structure formation. It is however not straightforward to link the insights gained by cosmological simulations to observations of the Universe. The region that is best studied and accessible observationally is the Local Universe, i.e.\ the Local Group and its immediate large-scale environment. A very attractive method is provided by the constrained realizations (CR) algorithm \citep{Hoffman1991}, which allows to impose observational data as constraints on the ICs. The simulations obtained from such constrained ICs are able to yield structures which can closely mimic those in the actual Local Universe. These \emph{constrained simulations} provide an ideal numerical laboratory to study various aspects of structure formation in the Local Universe, from the dynamics of the nearby massive superclusters down to the formation, evolution, and properties of the Local Group itself and its satellite galaxies.

The key requirement for running constrained simulations is to generate a good estimate of the Gaussian initial conditions. The constrained simulations method implements the Bayesian approach in order to estimate and reconstruct the most probable field given the assumed Gaussian PDF and some observational data, filtering out statistical noise from observational errors. 

A very useful algorithm is provided by the formalism of the Wiener Filter (WF, \citet{Wiener1949}, \citet{Zaroubi1995}). The WF reconstruction is then augmented with random small-scale perturbations using the Hoffman-Ribak algorithm \citep{Hoffman1991} to yield a constrained realization of ICs with the required statistical properties, i.e.\ enough power on all scales resolvable by the simulation. Two directions can be taken for getting appropriate input data: using either the redshift positions or the peculiar velocities of observed galaxies. Early attempts by \citet{Kolatt1996}, \citet{Bistolas1998}, and \citet{Mathis2002} used redshift catalogues to obtain an estimate of the underlying field, taking this field backwards into the linear (Gaussian) regime using the Eulerian 
Zeldovich-Bernoulli equation \citep{Nusser1992}, and then generating initial conditions for simulations with the CR algorithm. More recently, this approach was enhanced by \citet{Lavaux2010b} by using the MAK reconstruction \citep{Frisch2002,Brenier2003,Mohayaee2003}, and by \citet{Kitaura2012arXiv}, who performed the Gaussianization by Hamitonian sampling with a Gaussian-Poisson model.

\subsection{CLUES simulations from peculiar velocities}

Although the redshift space approach can already produce a reasonable estimate of the largest observed structures, using galaxy peculiar velocities for input data has several benefits. Catalogues of galaxy peculiar velocities (more precisely, their radial components) are constructed from galaxy redshift measurements and independent estimates of their distance (see e.g.\ \citealt{Tully2009}). Direct distance measurements do not suffer from redshift distortions, and the derived peculiar velocities provide a direct tracer of the underlying gravitational potential and therefore the total mass distribution without suffering from the galaxy bias. Statistically, the peculiar velocity field at $z=0$ is much closer to a Gaussian distribution than the density field traced by galaxy redshift positions, which facilitates Bayesian reconstruction. Moreover, peculiar velocities feature strong large-scale correlations and therefore allow to extrapolate the reconstructed field into regions not covered by measured galaxies, such as outside of the data zone, and inside the Zone of Avoidance which is obscured by the Milky Way disk; as a result, reconstructions using peculiar velocities are typically much less sensitive to data that is sparse, incomplete, and distributed in a statistically inhomogeneous way. This approach of using radial peculiar velocities for Bayesian reconstruction was pioneered by \citet{Zaroubi1999}, who laid out the necessary theoretical framework, and the first high-quality simulations based on it were conducted by \citep{Klypin2003}. Building upon this work, the CLUES project \citep{Gottloeber2010arXiv} is dedicated to construct simulations that reproduce the Local Universe and its key ingredients, such as the Local Supercluster (LSS), the Virgo cluster, the Coma cluster, the Great Attractor (GA) and the Perseus-Pisces supercluster. The observational data is provided by a collaboration with the Cosmicflows program \citep{Courtois2011a,Courtois2011b,Courtois2012arXiv,Tully2012arXiv}, whose goal is to determine galaxy distances for 30,000 galaxies in the Local Universe with systematic errors below 2\%. CLUES simulations are used to study a variety of aspects of the Local Universe. This includes the question whether our own Local Group is a ``typical'' object and studies of its mass accretion history \citep{Forero2011}, the unusually cold local Hubble low \citep{MartinezVaquero2009}, and the abundance and spatial distribution of satellite galaxies in the Local Group \citep{Libeskind2010,Klimentowski2010,Knebe2011a,Knebe2011b,DiCintio2012ArxivA}. 

CLUES simulations have also been applied to investigate the possibility of WDM models \citep{Yepes2009} and to simulate possible observations of dark matter \citep{Cuesta2011}. The tremendous increase in data quality of peculiar velocity observations in recent years presents a serious challenge for the simulators: the present methods relying on this data to generate constrained initial conditions and constrained simulations have significant limitations and need to be improved as well, in order to optimally utilize the wealth of additional information contained in the new data sets. This challenge provides the main motivation for this work.

The main drawback of current CLUES simulations is that, while they are able to reproduce massive clusters like the Virgo, the Great Attractor (GA), Coma, and Perseus-Pisces, they do not directly constrain the structure on scales smaller than such massive clusters, which are therefore dominated by the random component added by the CR algorithm (although all objects that emerge on these smaller scales are still located in the correct large-scale environment). The reason is that the CR formalism is formulated completely within the formalism of  Gaussian fields, thereby assuming that the linear theory of density perturbations is valid on all scales. The data used as input is however observed at present time, after having undergone the highly non-linear structure formation process, and the linearity assumption is only valid  down to a certain length scale. A particularly strong error source is the cosmic displacement field: the galaxies observed today are located at different positions than their progenitors in the linear regime due to their motion that arises from the large-scale gravitational field. At $z=0$, the average amplitude of this displacement field is $\approx 10$ Mpc/$h$, severely limiting the validity of the linear formalism on scales smaller than this. Accounting for this non-linearity must therefore be done by a reconstruction of the displacement field and the initial distribution of the data in the linear regime, given the final distribution. This task is commonly termed \emph{Lagrangian reconstruction} and is a highly non-trivial undertaking due to the underlying strongly non-linear dynamics. Methods developed so far usually aim at reconstructing galaxy orbits by minimizing the action \citep{Croft1997,Peebles2001,Frisch2002,Brenier2003,Mohayaee2003,Lavaux2010b}, but have until now exclusively focused on galaxy redshift surveys, and therefore their accuracy is limited by the high non-linearity and strong observational biases of such data \citep{Lavaux2008}; in addition, they are all very expensive computationally, and they do not make use of the known statistical properties of the initial conditions that are to be reconstructed, thus lacking self-consistency with the assumed cosmological model.

\subsection{Outline}

This work is the first part of a serie of three papers by Doumler et al. In this series, we present a novel self-consistent method for Lagrangian reconstruction that is designed for the application to peculiar velocity data, which we call Reverse Zeldovich Approximation (RZA). We test the validity of this method using mock data from a given cosmological simulation, serving as the test Universe. Applying the RZA does not require expensive computations, and at the same time allows to generate a good estimate of the cosmological initial conditions which can then be used to run constrained simulations. The resulting simulations have a significantly higher accuracy compared to those run from ICs generated with the previous method (without Lagrangian reconstruction). In this first paper, we present the method itself, how to apply it to realistic observational data, and how to use it to construct ICs for constrained simulations. 

This paper is organised as follows: In Section \ref{sec:rza}, we review the CR method, the problem of Lagrangian reconstruction, and introduce the idea of the RZA. In Section \ref{sec:rzaatz0}, we verify the validity of the RZA on a cosmological simulation. Section \ref{sec:mocks} describes how the RZA can be applied to realistic data. Here, we test its performance using mock data drawn from the same simulation, but with observational errors, a relatively small data volume, high incompleteness and knowledge of only the radial component of the haloes' peculiar velocities. We then describe in Section \ref{sec:ics} how the RZA can be used to reconstruct the cosmological initial conditions of the input data. Our findings are summarized and discussed in Section \ref{sec:summary}.

\section{Theoretical framework}
\label{sec:rza}

\subsection{The Wiener Filter}
\label{subsec:rza_wf}

Within the linear theory of Gaussian random fields, the Wiener Filter (WF) reconstruction method is an established tool for the purpose of reconstructing the underlying density ($\delta$) and peculiar velocity ($\bl v$) fields from sparse and noisy data sets \citep{Rybicki1992,Fisher1995,Zaroubi1995,Erdogdu2006,Hoffman2009,Kitaura2009,Courtois2012}.  In the case of peculiar velocity data, we are given the radial component $v_r$ of the three-dimensional peculiar velocity field $\bl v$ sampled at some discrete positions $\bl r$. This data is additionally corrupted with observational errors $\varepsilon$, so that the input consists of a set of datapoints or ``constraints'', $\Gamma = \{ C_i \}$ with $C_i = c_i + \varepsilon_i$, where $c_i=v_{r,i}$ . 
Here, a data point is separated into the actual signal that is contributed by the underlying error ($c_i$) and the statistical observational error ($\varepsilon$). Two basic underlying assumptions are made here, one is that the errors are statistical and all the systematic errors have been accounted for and that the errors and the underlying dynamical fields are statistically independent \citep{Zaroubi1995}.
For this kind of data, the Wiener Filter is used in its cartesian real space formulation; for a full description of this formalism see \citet{Zaroubi1999}. The Wiener Filter mean field is given by 
\begin{align}
\label{eq:deltawf}f
\delta^\tm{WF} (\bl r) = \langle \delta(\bl r) c_i \rangle \,\langle C_i C_j \rangle^{-1}\, C_j \;\;\;, \\
\label{eq:vwf}
\bl v^\tm{WF} (\bl r) = \langle \bl v(\bl r) c_i \rangle \,\langle C_i C_j \rangle^{-1}\, C_j \;\;\;,
\end{align}
where $\langle C_i C_j \rangle$ is the data autocorrelation matrix of the data, and $\langle \delta(\bl x) c_i \rangle$ and $\langle \bl u(\bl x) c_i \rangle$ are the cross-correlation matrices  of the data with the fields to be reconstructed. 
One should note that the cross-correlation matrices do not correlate the noise, i.e. the statistical errors, with the signal and therefore the cross-correlation with $c_i$ is written. The data auto-correlation matrix is written:
\begin{equation}
\label{eq:autocorrelation}
\langle C_i C_j \rangle =  \langle c_i c_j \rangle + \sigma{_i^2} \delta_{ij},   \;\;\;,
\end{equation}
where
\begin{equation}
\label{eq:error-matrix}
 \langle \varepsilon_i \varepsilon_j \rangle = \sigma{_i^2} \delta_{ij}.  \;\;\;,
\end{equation}
Equation \ref{eq:error-matrix} assumes that the errors of different data points are statistically independent. The error matrix needs to be tailored to the data base at hand.

Because we are dealing with Gaussian random fields, the  correlation matrices which describe the underlying fields  are completely defined by the power spectrum $P(k)$ of the prior model. It can be shown that the WF mean field is the optimal estimator for Gaussian random fields. It is equivalent to the minimal variance estimator and the Bayesian conditional mean field (most probable field given the data and the prior model). The data $c_i$ describes only one component $v_\mu = \hat{\bl e}_\mu \cdot \bl v$, where $ \hat{\bl e}_\mu = (\bl r - \bl r_0) / |\bl r - \bl r_0|$ is the unit vector along the radial direction with respect to the observer position $\bl r_0$. This requires to compute the correlation functions $\langle \delta v_\mu \rangle$  and $\langle v_\mu v_\nu \rangle$  for arbitrary $\hat{\bl e}_\mu$, $\hat{\bl e}_\nu$, which is given by
\begin{align}
\label{eq:mucomp}
\langle \delta (\bl r^\prime) v_\mu (\bl r^\prime + \bl r)\rangle = \hat{\bl e}_\mu \cdot \langle \delta(\bl r^\prime) \    \bl v(\bl r^\prime + \bl r) \rangle \;\;\;,  \\
\label{eq:munucomp}
\langle v_\mu  (\bl r^\prime) v_\nu (\bl r^\prime + \bl r)\rangle = \hat{\bl e}_\mu \cdot \langle \bl v(\bl r^\prime)   \  \bl v(\bl r^\prime + \bl r) \rangle \cdot \hat{\bl e}_\nu \;\;\;, 
\end{align}
where the density-velocity correlation vector is given by
\begin{align}
\label{eq:zeta}
\langle \delta(\bl r^\prime) \, \bl v(\bl r^\prime + \bl r) \rangle_\alpha = \frac{\dot a f}{(2 \pi)^3} \int_0^\infty \left( \frac{-ik_\alpha}{k^2} \right) P(\bl k) e^{-i \bl k \cdot \bl r} \tm{d}\bl k   \;\;\;,
\end{align}
and the velocity-velocity correlation tensor is given by
\begin{align}
\label{eq:psi}
\langle \bl v(\bl r^\prime)  \,\bl v(\bl r^\prime + \bl r) \rangle_{\alpha \beta} = \frac{(\dot a f)^2}{(2 \pi)^3} \int_0^\infty \left( \frac{k_\alpha k_\beta}{k^4} \right) P(\bl k) e^{-i \bl k \cdot \bl r} \tm{d}\bl k   \;\;\;,
\end{align}
where $\alpha,\beta \in \{x,y,z\}$ are the three cartesian components, and $P(\bl k)$ is the cosmological power spectrum of the assumed prior model. See \citet{Zaroubi1999} for more details on how to evaluate expressions (\ref{eq:zeta}) and (\ref{eq:psi}).

The formalism presented here does not account for the fact that the observed peculiar velocities, used here as constraints, sample a non-linear velocity field. Peculiar velocities of galaxies outside the cores of rich clusters  constitute a good proxy to the linear velocity field  in the standard model of cosmology. We adopt here the   \citet{Bistolas1998} partial remedy for their quasi-linearity. This consists of adding a constant term to the diagonal of the error matrix, aimed at making the data statistical compatible with the prior model. In the case where the assumed prior model is  'truly' the model of the universe, that the data does not deviate from linear theory and that the errors are estimated correctly, then the reduced chi-squared of the data, $\tilde{\chi}^2$, should be very close to unity, where
\begin{equation}
\label{eq:chisq}
\tilde{\chi}^2 = { C_i  \, \langle  C_i C_j \rangle^{-1} \  C_j \over M} 
\end{equation}
and $M$ is the number of data points. In all observational data bases used before within the CLUES collaboration and the mock catalogs constructed here the resulting reduced chi-squared has always been found to be larger than unity, indicating an excess of power over what is predicted by the prior model within the linear regime. Following  \citet{Bistolas1998} we augment the error matrix with a constant $\sigma_{NL}$ term,
\begin{equation}
\label{eq:sigmanl}
 \langle  \varepsilon_i  \varepsilon_j  \rangle = (\sigma{_i^2}  \  + \   \sigma{_{NL}^2}  )\  \delta_{ij}.  \;\;\;,
\end{equation}
The value of $\sigma_{NL}$ is tuned so as to equate $\tilde{\chi}^2=1.0$.

As a part of this work, we developed the numerical software package \textsc{ICeCoRe}\footnote{this acronym stands for ``Initial Conditions \& Constrained Realizations''.}, a highly efficient parallelized code written in C++. 
One part of \textsc{ICeCoRe}'s functionality is the ability to efficiently compute the WF mean field for very large datasets. This includes correctly inverting the $M \times M$ data auto correlation matrix (where $M = |\Gamma|$ is the number of datapoints) and then evaluating the WF mean field on a full three-dimensional cubic grid with high resolution. 
The matrix inversion is performed by using the Cholesky decomposition method, whose main advantage is its high numerical stability. 
The computational cost of the inversion scales with $\mathcal{O}(M^3)$, and the maximum rank $M$ of the matrix that can be inverted is limited in principle only by the available memory. For $M = 4000$, the inversion takes only a couple of seconds in serial. For matrices larger than $M \approx \  (1\ - \ 2)\times 10^4$, it is recommended to switch to parallel machines. For $M = 5\times 10^4$, the inversion takes around 60 minutes with eight OpenMP threads on two quad-core 2.4 GHz Intel Xeon processors, with 18 GB of memory required for the inversion. 
We successfully tested \textsc{ICeCoRe} for up to $M = 10^5$ data points and have not encountered any issues of numerical stability.  \textsc{ICeCoRe} also contains a number of utilities to manipulate the input data, to vary the implementation details of the WF algorithm, and to post-process the obtained three-dimensional fields.

\subsection{The Constrained Realizations algorithm}
\label{subsec:rza_cr}

At the heart of the CR method (in its linear-theory form) lies the idea to use the Gaussian field reconstructed with the WF as an estimate of the cosmological initial conditions underlying the data. However, while the WF is the optimal estimator for Gaussian random fields, it is also a very conservative estimator. It tends towards the unconstrained mean field (i.e., the null field) in regions not sampled by data and if the data is strongly corrupted by noise. It is therefore not power-preserving: the solution will lack power on scales not covered by the constraints. However, the initial conditions required to run a cosmological simulation must be a Gaussian random field realization of the assumed power spectrum $P(k)$ at all scales resolved by the simulation. The CR method \citep{Bertschinger1987,Binney1991,Hoffman1991,Hoffman1992,Weygaert1996,Prunet2008} provides a way to compensate for the missing power by adding 
fluctuations 
from an independently generated random realization (RR). The optimal exact algorithm for generating a CR was discovered by \citet{Hoffman1991}. 
We first generate a random realization, $\delta^\tm{RR}$, and then ``observe'' it to get a set of ``mock constraints'' $\tilde\Gamma = \{ \tilde C_i \}$. The $\tilde C_i$ constrain the same quantities at the same positions as the original data $C_i$, but the values are drawn from $\delta^\tm{RR}$ instead. Then, a constrained realization is generated with
\begin{align}
\label{eq:deltacr}
\delta^\tm{CR} (\bl r) = \delta^\tm{RR} (\bl r) + \langle \delta(\bl r) c_i \rangle \,\langle C_i C_j \rangle^{-1}\, (C_j - \tilde{C}_j) \;\;\;, \\
\label{eq:vcr}
\bl v^\tm{CR} (\bl r) = \bl v^\tm{RR} (\bl r) + \langle \bl v(\bl r) c_i \rangle \,\langle C_i C_j \rangle^{-1}\, (C_j - \tilde{C}_j) \;\;\;.
\end{align}
In regions where the data $c_i$ are dense and accurate, such a CR will be dominated by the data and tend towards the WF mean field, i.e. the most probable linear field given the data and the $P(k)$. Conversely, the result will be dominated by the random component $\delta^\tm{RR}$ in regions not sufficiently constrained by the $c_i$, i.e. where they are sparse, noisy, or not present at all. In general, the result will show a smooth transition between both regimes and produce a Gaussian random field that is statistically homogeneous, obeys the assumed prior model $P(k)$, and can therefore be used for cosmological ICs. An efficient implementation of this CR algorithm is provided in our \textsc{ICeCoRe} code. In order to start an $N$-body simulation from $\delta^\tm{CR}$, we merely have to linearly scale it to the chosen starting redshift $z_\tm{init}$ and to sample it with $N$ discrete particles (for example with the established Zeldovich method, see \citealt{Efstathiou1985}). This is the basic method how CLUES simulations have been set up until now, aside from some practical implementation issues (see \citealt{Gottloeber2010arXiv}).

Since the WF operator is mathematically very similar to the CR operator, except that the latter adds the RR terms, in the following we will refer to the general algorithm as WF/CR.

\subsection{The Zeldovich approximation and Lagrangian reconstruction}

The linear theory of structure formation describes the density and peculiar velocity fields as time-independent fields except for the cosmology-dependent growth factor $D$ and growth rate $f$ = $\tm{d ln}(D)/\tm{d ln}(a)$, considering only the growing mode,
\begin{align}
\label{eq:lineartheory}
\delta(\bl x,z) &= D(z) \; \delta_0(\bl x) \;\;\;, \\
\bl u(\bl x,z) &= - \dot a f \; \bl \nabla^{-1} \delta(\bl x,z) \;\;\;,
\end{align}
where we use comoving coordinates $\bl x = \bl r /a$, $\bl u = \tm{d}\bl x/\tm{d}t/a $. Moving beyond the linear theory of density perturbations, the time evolution of the cosmic density distribution can be described in the formalism of Lagrangian perturbation theory. If the initial conditions are sampled by some tracers at initial (homogeneously distributed) Lagrangian positions $\bl q$, then these positions will change as a result of motions induced by the cosmic gravitational field, and the actual positions $\bl x(z)$ at any redshift $z$ are given by
\begin{equation}
\label{eq:lagrange}
\bl x (z) = \bl q (\bl x) + \bl \psi (\bl x,\  z) \;\;\;,
\end{equation}
and as a result the density and velocity fields will change as well. The deviation of these fields from the initial conditions as they evolve forward in time can therefore be described in terms of the displacement field $\bl \psi(z)$. Here, the first-order approximation has proven to be very useful and surprisingly accurate well into the quasi-linear regime, namely that the displacement field is itself constant except for the growth factor $D$ and proportional to the peculiar velocity,
\begin{align}
\label{eq:basiczeldovich}
\bl \psi (\bl x, z) &= D(z) \; \bl \psi_0 (\bl x) = - D(z) \; \nabla^{-1}  \; \delta_0(\bl x)  \;\;\;, \\
\bl u (\bl x, z) &= \dot a f \bl \psi(\bl x, z) \;\;\;.
\end{align}
This is the famous Zeldovich approximation \citep{Zeldovich1970,Shandarin1989}. Considering the next term, i.e.\ the acceleration, leads to second-order Lagrangian perturbation theory or 2lpt \citep{Buchert1994,Bouchet1995}; this can be further expanded to higher orders.

While Lagrangian perturbation theory is doing well in describing the evolution forward in time until non-linearities set in, the situation becomes much more complicated if we want to do the reverse, i.e.\ reconstruct the initial field $\delta_0(\bl q)$ from a sampling of the evolved density or velocity field at $z=0$. This is equivalent to tracing the field back in time from $z = 0$ to some $z_\tm{init}$ where $\delta(\bl x) \approx \delta(\bl q)$, at which point we would have obtained the cosmological initial conditions we are looking for. In other words, we need a mapping of Eulerian coordinates $\bl x = \bl r$ at $z=0$ to the corresponding Lagrangian coordinates $\bl q$, that is, we need to obtain the cosmological displacement field $\bl \psi$ in Eq. (\ref{eq:lagrange}) at $z=0$. This is an extremely complex task because cosmic structure formation and evolution is a highly non-linear process. Even if we could reconstruct $\delta(\bl r)$ at $z=0$ to arbitrary precision (which is of course impossible), we could not just integrate the equations of motion back in time, such as by running a cosmological $N$-body simulation backwards, to recover $\delta(\bl q)$ at $z_\tm{init}$. Although gravity is in principle invariant under time reversal, reversing the time direction would turn the decaying mode of perturbation growth into a growing mode, which will rapidly increase and amplify any uncertainties in the data or even slight numerical errors until they eventually dominate the solution. With such a procedure, the probability of recovering the highly ordered state of homogeneous and almost uniform initial conditions will be infinitely small. Additionally, such a time reversal could not be carried out in a consistent way, because shell crossing and dissipational processes effectively erase information about the initial state of the Universe \citep{Crocce2006a}.
Yet, it  should be  mentioned that forward modeling algorithms should be able to overcome  these problems   \citep[e.g.][]{Kitaura2010,2012arXiv1203.3639J}.

\subsection{Lagrangian reconstruction from peculiar velocities}

Previous simulations conducted in the CLUES framework \citep{Gottloeber2010arXiv} directly used observational peculiar velocity data as constraints for the CR algorithm to obtain the constrained initial conditions. The main drawback of this method is that it completely neglects the fact that the data does \emph{not} trace the initial conditions at some high initial redshift $z_\tm{init}$, where all resolved scales lie in the linear regime of density perturbations, but instead the non-linearly evolved field at $z=0$. Therefore, the accuracy of the method is limited to large scales above some critical scale where the difference between the initial and the final field becomes non-negligible. 

\begin{figure*}
\centering
\includegraphics[scale=1]{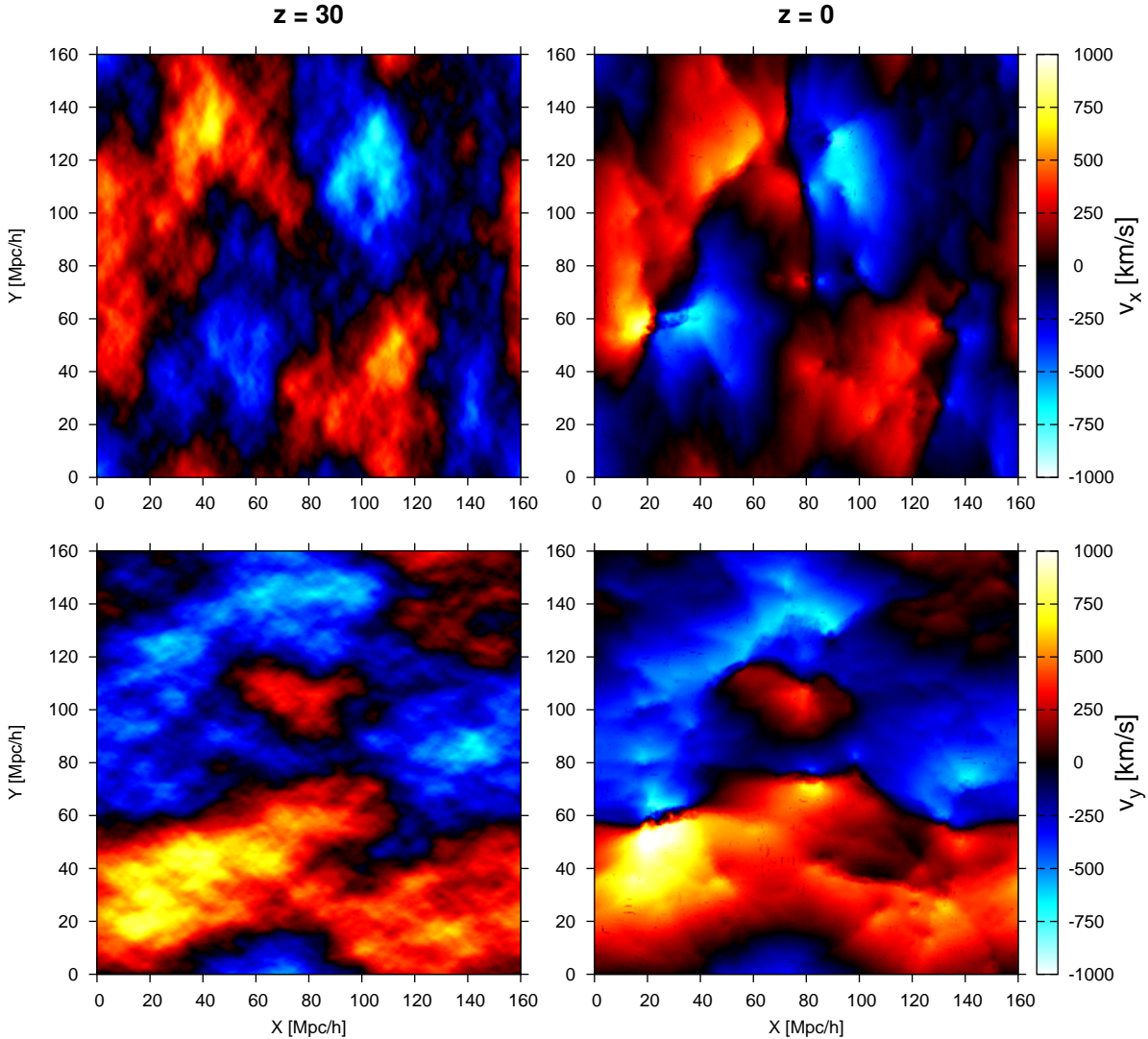}
\caption{Slice through the BOX160 simulation with boxsize 160 Mpc/$h$, slice thickness 10 Mpc/$h$ at the initial redshift $z=30$ (left) and the final redshift $z=0$ (right). Shown are projections of the two components of the peculiar velocity field that lie in the depicted plane, $v_x$ (top) and $v_y$ (bottom). The left panels were generated directly from the initial conditions computed on a grid; right panels were generated by a binning of the evolved particle velocities to a regular $256^3$ grid with a TSC kernel. The z=30 velocity field (left) is scaled to z=0 by the linear theory.}
\label{fig:simuvfields_paper}
\end{figure*}

For the initial vs. evolved \emph{peculiar velocity} field, this critical scale is around $\approx 10$ Mpc/$h$. It is an interesting fact that this critical scale is much lower than the one for the initial vs. evolved \emph{density} field, where at $z=0$ non-linearity, or in other words non-Gaussianity, is noticeable at much larger scales. Figure \ref{fig:simuvfields_paper} shows the peculiar velocity field component-wise at $z_\tm{init} = 30$ (left) and at $z=0$ (right) for a cosmological simulation. Both look remarkably similar. The evolved field is shifted by the cosmological displacement field $\bl \psi$, which varies locally and has an average amplitude of $\approx 10$ Mpc/$h$, and there is a non-linear enhancement of the velocity amplitude at even smaller scales due to the gravitational collapse of structure and another non-linear component due to virial motions. However, on scales above those of gravitationally bound objects (i.e.\ a few Mpc/$h$) the velocity field $\bl v (\bl r)$ is very close to the Gaussian distribution expected by linear theory \citep{Sheth2001b,Hamana2003}. Because of these properties, the peculiar velocity approach works reasonably well for the large-scale structure even without any ``time machine'' method at all, as seen in previous CLUES simulations. The scale down to which this works will be determined by the scale of $\bl \psi$. On the other hand, any attempt to construct a constrained realization from tracers of the density field must include some Gaussianization procedure, which can be Eulerian  (such as the methods of \citealt{Weinberg1992,Nusser1992}) or also include a Lagrangian reconstruction (such as the already mentioned action-minimizing methods). But avoiding the Gaussianization problem by using peculiar velocities still leaves us with a significant error due to the displacement field. This means that the constraints we use for Eqs. (\ref{eq:deltacr}) and (\ref{eq:vcr}) are displaced by $\bl \psi$ relative to the true initial conditions. As a result, the cluster positions in the evolved constrained simulations will be off by $\approx 10$ Mpc/$h$ on average (and more in regions where the displacement field amplitude is higher) compared to the observed configuration at $z=0$ \citep{Gottloeber2010arXiv}.

It is clear that, despite the favourable properties of peculiar velocities, in order to increase the quality of our initial conditions and constrained simulations down to scales lower than the amplitude of $\bl \psi$ and to remove the significant systematic position errors caused by $\bl \psi$, we need a Lagrangian reconstruction scheme that can be applied directly to the peculiar velocity data. In this work, we develop such a method, the basic idea of which will be described in the next section.

\subsection{The Reverse Zeldovich Approximation}

In the remainder of this paper, we will denote the peculiar velocity $\bl v$ and the physical position $\bl r$ as the special case $z=0$ of the comoving velocity $\bl u(z)$ and the comoving position $\bl x(z)$, respectively. In the following, we consider again the Zeldovich approximation. Then, at $z=0$, the positions $\bl r$ and peculiar velocities $\bl v$ should be given as:
\begin{align}
\label{eq:rza_zeldovichz0r}
\bl r &= \bl q + \bl \psi(\bl q) \;\;\;, \\
\label{eq:rza_zeldovichz0v}
\bl v(\bl r) &= H_0 f\bl \psi(\bl q) \;\;\;,
\end{align}
with $\dot a = H_0$ at $z=0$. This is, of course, a very crude assumption: by $z=0$ the approximation will, in general, break down in overdense regions due to shell crossing and nonlinear gravitational interaction. However, we assume its validity for the time being. Thinking in the other direction, this enables us to obtain an estimate of the displacement field $\bl \psi$ and the Lagrangian position $\bl q$, if the velocity $\bl v$ is known. By design, we are not interested in recovering the full displacement field $\bl \psi(\bl r)$ everywhere in the computational box; rather, it is enough to recover it at the positions of the available discrete data points used as the input data $c_i$. We can then use these discrete displacement field values and their reconstructed Lagrangian positions to constrain initial conditions, utilising the CR algorithm. This approach works well with the first-order Zeldovich approximation, since it is completely local and can be performed at discrete locations, and with sparse and inhomogeneously sampled input data, whereas higher-order extensions depend on integrals over the computational volume.

Let us consider a single position $\bl r$ where we have given a discrete value of the peculiar velocity field $\bl {v}(r)$ at $z=0$ (we will deal later with observational errors and the fact that only the radial part $v_r(r)$ is actually available). In what we call the Reverse Zeldovich Approximation (RZA), we can now simply reverse Eqs. (\ref{eq:rza_zeldovichz0r}) and (\ref{eq:rza_zeldovichz0v}) and estimate the displacement of this data point as
\begin{align}
\label{eq:psi_rza}
\bl \psi^{\tm{RZA}} = \frac{\bl v}{H_0 f}\;\;\;,
\end{align}
and the initial position of the halo by
\begin{align}
\label{eq:xinit_rza}
\bl x_{\tm{init}}^{\tm{RZA}}  = \bl r - \frac{\bl v}{H_0 f}\;\;\;,
\end{align} 
where $\dot a f = H_0 f$ at $z=0$, and we approximated $\bl x_{\tm{init}}^{\tm{RZA}} \approx \bl q^{\tm{RZA}}$, since $z_\tm{init}$ will be chosen sufficiently high so that there $\bl x \approx \bl q$. Both $ \bl x_{\tm{init}}^{\tm{RZA}}$ and $\bl \psi^{\tm{RZA}}$ could then be used to place a constraint for generating initial conditions at $z_\tm{init}$ (see Section \ref{sec:ics}). This reconstructed initial position will be, in general, at some distance $d^{\tm{RZA}}$ from the actual initial position, which we define to be the ``RZA error'' $d^\tm{RZA}$ for this data point,
\begin{align}
\label{eq:rza_error}
d^{\tm{RZA}} = \left|\bl x_{\tm{init}} - \bl x_{\tm{init}}^{\tm{RZA}}\right| = \left|\bl x_{\tm{init}} - \bl r + \frac{\bl v}{H_0 f}\right| \;\;\;,
\end{align}
where $\bl x_\tm{init}$ is the actual true position of this data point in the initial conditions at $z_\tm{init}$. It is interesting to study how well this relatively simple approach performs in practice. The RZA error $d^{\tm{RZA}}$ provides a simple one-dimensional quantity useful to quantify the scale length down to which the RZA is valid for different data points and  environments where they are located. In the next section, we do so using a cosmological simulation, where both the initial conditions and the distribution at $z=0$ are known. 

In Sections  \ref{sec:mocks}  and \ref{sec:ics} we will then use the reconstructed $\bl \psi^{\tm{RZA}}$ and  $\bl x_{\tm{init}}^{\tm{RZA}}$ to obtain a new set of constraints for constraining initial conditions with the CR algorithm, which are more suitable than the original $c_i$ due to the Lagrangian reconstruction.

We have to note here that there have been similar attempts to estimate $\bl \psi$ by reversing the Zeldovich approximation, but from the \emph{density} field at $z=0$, using the linear-theory assumption $\delta = - \bl \nabla \cdot \bl \psi$ \citep{Eisenstein2007,Noh2009} or higher-order lpt \citep{Falck2012}. The quality of these reconstructions is unfortunately far too poor to obtain a set of constraints usable for constrained simulations, because of the high non-linearity of $\delta$ at $z=0$. However, we find that due to their nature, the application of this idea to peculiar velocities, i.e.\ our RZA method, performs substantially better.


\section{Validity of the RZA at $z=0$}
\label{sec:rzaatz0}

\subsection{The test simulation}

For this study, we use the BOX160 simulation performed within the CLUES project. This is a constrained simulation of the Local Universe with a boxsize of 160 Mpc/$h$, set up with the WMAP3 cosmological parameters\footnote{Although these parameters are outdated from today's perspective, we have found in other (non-constrained) test simulations that switching to a more recent set of parameters like WMAP7 does not influence the results obtained from RZA reconstruction.} $\Omega_m = 0.24$, $\Omega_\Lambda = 0.76$, and $\sigma_8 = 0.75$ and a matching $\Lambda$CDM power spectrum. The simulation contains a large-scale structure resembling the observed Local Universe, with objects corresponding to the Virgo, Coma, Perseus-Pisces, and Hydra-Norma-Centaurus (Great Attractor) clusters, and a Local Group candidate. Details of the simulation are discussed in \citet{Cuesta2011}. Theoretically, investigating the validity of the RZA method could be done on any (non-constrained) cosmological simulation. However, we choose a constrained CLUES simulation on purpose, so that the gained insight can easier be applied to the observed Local Universe. 

By employing a cosmological simulation as the model universe, the relationship between peculiar velocities and cosmological initial conditions can be studied directly, because we have complete access to both the initial conditions and the distribution at $z=0$. We use peculiar velocities of dark matter haloes at $z=0$ as a proxy for observable galaxy peculiar velocities. In the current theoretical picture of galaxy formation, all galaxies reside inside the potential wells of dark matter haloes. It is therefore a reasonable assumption that the observed galaxy peculiar velocities follow the peculiar velocities of their surrounding dark matter haloes. These are in turn directly accessible in the simulation snapshots. The BOX160 is a collisionless-matter-only simulation. This prevents us from a proper modeling of the bias induced by the fact that galaxies are used to sample the velocity field. None of the current galaxy formation simulations has a dynamical range large enough that enables the resolution of sub-galactic scales with large cosmological boxes. Hence, one needs to resort to DM-only simulations for modelling the galaxy distribution.
 
We extract a halo catalogue from the BOX160 simulation using AHF (Amiga's halo finder; \citealt{Knollmann2009}) at the $z=0$ snapshot of the evolved simulation. The peculiar velocities of each halo are extracted by AHF and defined as the average velocity vector of all dark matter particles within a halo's virial radius $R_\tm{vir}$. Only haloes of mass $\log (M/M_\odot) \geq 11.5$ from the simulation are considered; we want to discard haloes that are either too poorly resolved to obtain reasonable estimates on their peculiar velocity, or too small to host galaxies that would be observable in a galaxy distance survey. In the following, we investigate how well the displacement field $\bl \psi$ with respect to the simulation's initial conditions can be recovered from these halo velocities. Observational data features only the radial component of the peculiar velocities, and suffers from significant sparseness, incompleteness, a very limited data volume and observational errors. We will investigate the impact of these limitations later in Section \ref{sec:mocks}.

\subsection{RZA reconstruction}

\begin{figure}
\centering
\includegraphics[trim=0.9cm 0cm 1.1cm 0cm, clip=true, scale=0.93]{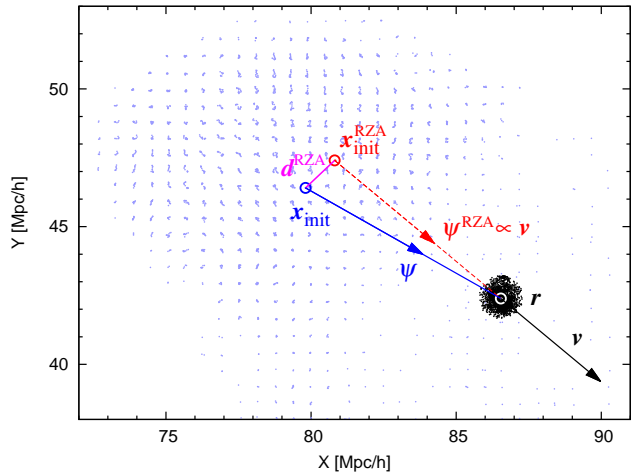}
\caption{RZA on a simulated dark matter halo identified at $z=0$ with virial mass $M=7.9 \times 10^{12} M_\odot$, virial radius $R_{\tm{vir}} = 410$ kpc/$h$. Black dots: positions of all particles inside $R_{vir}$ at $z=0$, with mean velocity $\bl v$ and the halo centre at position $\bl r$. Blue dots: positions of the same particles in the initial conditions at $z_{\tm{init}}=30$, with centre-of-mass at $\bl x_{\tm{init}}$ (``initial position''). $\bl \psi$ (blue arrow): actual displacement, with $|\bl \psi|=8.36$ Mpc/$h$. $\bl \psi^{\tm{RZA}}$ (red arrow): RZA reconstructed displacement. $\bl x_{\tm{init}}^{\tm{RZA}}$: RZA reconstructed initial position. $d^{\tm{RZA}}$ (purple): the RZA error.}
\label{fig:protohalo}
\end{figure}

Figure \ref{fig:protohalo} illustrates a typical medium-size halo with virial mass $M=7.9 \times 10^{12} M_\odot$ from the BOX160 simulation. The positions of particles within the virial radius $R_{\tm{vir}}$ that have been identified by AHF are shown with black dots. The halo has a position $\bl r$, which is defined by AHF as the position of the most bound particle, and a velocity $\bl v$, which is the mean velocity of all of the halo's particles. In the initial conditions at $z_{\tm{init}}=30$, the same particles, identified by their IDs, occupy the positions marked by the blue dots. Most of them form a coherent patch in space, the protohalo\footnote{Note that a small fraction of the particles at $z_{\tm{init}}$ is not connected to the protohalo patch, i.e.\ the Lagrangian volume is disjoint. Those particles mostly ended up in the virialised halo after accretion or merging processes and subsequent relaxation that happened much later than $z_{\tm{init}}$ during the non-linear structure formation process. We will not attempt to track this process for each individual halo and choose to not treat those particles separately, although they may affect the estimation of $\bl x_{\tm{init}}$ and $\bl \psi$. We also add that while the halo in Figure \ref{fig:protohalo} is the most common case, there are some more extreme cases with more disjoint Lagrangian regions; the corresponding haloes at $z=0$ are mostly the result of violent major merger events. For a thorough study of protohaloes, see \citet{Ludlow2011}.}. This patch covers an overdense region in the initial conditions that will later collapse to form the halo. If we neglect the tiny initial displacements of the particles $\bl \psi_{\tm{init}}$ at the initial conditions by approximating $\bl x_{\tm{init}} \approx \bl q$ for each particle, then the volume $V_{\tm{init}}$ occupied by the protohalo corresponds to the Langrangian volume of the halo. This volume can be relatively big, measuring about 10 Mpc/$h$ in diameter for this halo mass, or even more than 20 Mpc/$h$ for a massive cluster. Since the initial density distribution is almost uniform, this volume depends directly on the mass via $V_{\tm{init}} = M^3 / \bar \varrho$, where $\bar \varrho = 3 H^2 \Omega_\tm{m} / 8 \pi G$ is the mean cosmic density. We then define the ``initial position'' of the halo $\bl x_{\tm{init}}$ as the centre of mass of all its particles at initial redshift $z_\tm{init}$. We further define the displacement of the halo as $\bl \psi = \bl r - \bl x_{\tm{init}}$. For the following, it is important to remember that those are now quantities averaged over all particles that belong to the halo, and not values of the continuous fields at discrete points in space.

We assume that each of the selected haloes in the simulation hosts a galaxy with an observable peculiar velocity $\bl v$. We continue to use the assumption that the peculiar velocity of any observed galaxy follows the mean peculiar velocity of its surrounding dark matter halo host. Then we can directly use the velocities $\bl {v}(r)$ of the haloes in a simulation as a simple model for observational data. For each halo we now apply the RZA, i.e.\ Eqs. (\ref{eq:psi_rza}) and (\ref{eq:xinit_rza}), to produce our estimate of the displacement field, $\bl \psi^\tm{RZA}$, and the initial position of the halo $\bl x_\tm{init}^\tm{RZA}$. In the simulation, we can identify all particles of all haloes by their ID and find them in the initial conditions at $z_\tm{init}$, so that the true values $\bl \psi$ and $\bl x_\tm{init}$ are also known for each halo. With this we also determine the values of the RZA error $d^\tm{RZA}$ (Eq. \ref{eq:rza_error}) for each halo.

\subsection{The RZA displacement field}

\begin{figure}
\centering
\includegraphics[scale=1.0]{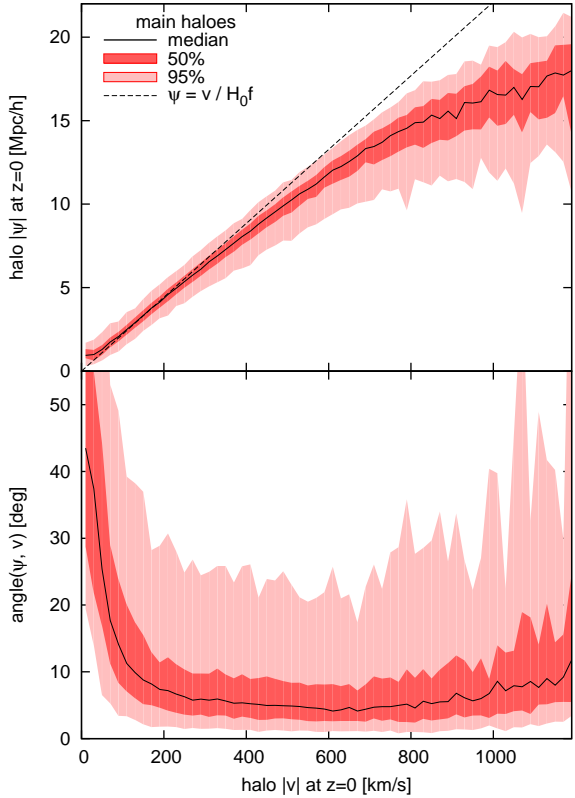}
\caption{Absolute displacement $|\bl \psi|$ vs. absolute peculiar velocity $|\bl v|$ (top row) and the angle between these two vectors (bottom row) at $z=0$ for main haloes, i.e.\ haloes after all their subhaloes (if any) have been grouped together with the main object.}
\label{fig:zeldovich_paper}
\end{figure}

\begin{figure}
\centering
\includegraphics[scale=1.0]{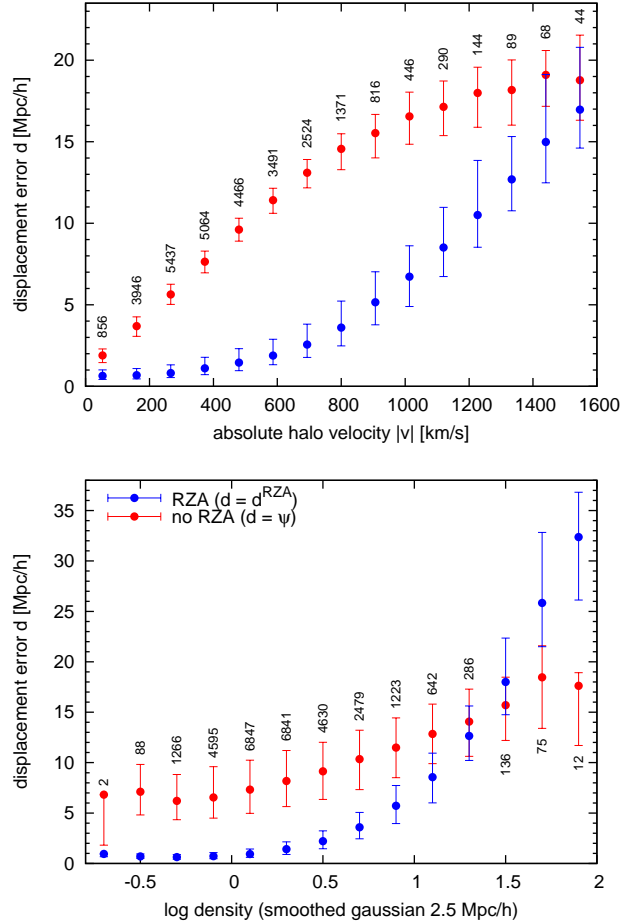}
\caption{Displacement error with RZA reconstruction ($d^{\tm{RZA}}$, blue) and without RZA reconstruction ($|\bl \psi|$, red), for all main haloes with mass $M > 10^{11.5} M_\odot$ inside BOX160, depending on the the absolute halo velocity (top) and the underlying local density (bottom). For each bin, the point is placed at the median, and the bar shows the interval between the $25^{\tm{th}}$ and $75^{\tm{th}}$ percentile. Additionally, the number of haloes in each bin is given. Note the different scales for the $d$ axis.}
\label{fig:rzaerrors_paper}
\end{figure}

The Zeldovich approximation breaks down on small scales where shell crossing occurs, marking the transition to the non-linear phase of structure formation. We expect that the RZA is not valid at all for haloes gravitationally bound to more massive haloes, i.e.\ orbiting, infalling, and merging substructure, since the magnitude and direction of their velocities at $z=0$ have been significantly altered by those processes. So in order to obtain a good estimate of the displacement field, those objects should be discarded. Physically, this would mean to detect and remove haloes that are gravitationally bound to more massive host objects, as well as ongoing major mergers and possibly other scenarios. This is a highly complicated task in itself, so we use a simplified scheme instead, which does not consider the underlying physics but works well for our purpose. Rather than properly finding actual subhaloes (such as in e.g.\ \citealt{Knollmann2009}), we simply detect haloes that share at least one dark matter particle with another halo and therefore presumably interact in some way or another, without investigating the nature of this interaction. We then call a halo a ``subhalo'', if it shares at least one dark matter particle with another halo more massive than itself. Conversely, we call a halo a ``main halo'' if it shares only particles with less massive haloes or with no other haloes at all. By this definition, the sphere described by the virial radius $R_\tm{vir}$ of a ``main halo'' will never overlap with the $R_\tm{vir}$ sphere of a more massive halo. In this way, we divide all the identified haloes in the simulation into two groups, subhaloes and main haloes, and keep only main haloes for the RZA analysis. This simple and rather conservative scheme is very effective in filtering out virial motions from the peculiar velocity data, since from clusters or other gravitationally bound systems only the main object will be retained\footnote{It is currently not yet clear how to construct a similarly effective grouping method for data points in observational galaxy peculiar velocity surveys. This problem is a subject of ongoing research in the CLUES project \citep{Courtois2012AN}.}.

Figure \ref{fig:zeldovich_paper} shows the absolute halo displacement $|\bl \psi|$ over the halo velocity $|\bl v|$ and the angle between those vectors at $z=0$ for the main haloes in BOX160. Subhaloes contain little to no information about their cosmological displacement from their momentary velocities at $z=0$. On the other hand, for the main haloes, the approximation holds reasonably well. In the low-velocity regime, which mostly corresponds to the low-overdensity regime, the relation is satisfied nearly perfectly; further up the plots reproduce the well-known tendency of the Zeldovich approximation to underestimate the total velocities, since it neglects the additional gravitational acceleration and deflection created by the dynamically changing density distribution. 
Nevertheless, the direction of displacement is in general conserved very well in the velocity vector: for the majority of identified haloes the angle between the two, $\alpha=\tm{acos}\left[(\bl v \cdot \bl \psi)\,/\,(|\bl v| \cdot |\bl \psi|)\right]$, lies below $\approx10^\circ$, and for 95\% of the main haloes, it lies below $\approx 30^\circ$. This could be further reduced by a more conservative ``grouping'' algorithm. We found that if we discard all haloes that are closer than 2.5 Mpc/$h$ to a more massive halo, then for half of the haloes the angle lies below $\approx 7^\circ$, and for 95\% of them below  $\approx 15^\circ$. After conducting this study, we learned that a similar investigation was already carried out by \citet{Sheth2001b}, who compared the peculiar velocities of haloes at $z=0$ with their \emph{initial} velocities $\bl u_0$ in the linear regime, instead of their displacement $\bl \psi$ as we did here. They likewise found that halo peculiar velocities at $z=0$ retain the information from the initial conditions very well, with deviations in the angle of motion of typically only $10^\circ$, if one thoroughly filters out virial motions. Our results are in very good agreement with theirs. 

The remainder of this analysis concentrates on the main haloes, since we now established that subhaloes should not be considered for RZA. This leaves us with a total of 29122 objects with $\log (M/M_\odot) \geq 11.5$ within the BOX160. We focus on the error $d^{\tm{RZA}}$ of the RZA on the initial halo position guess (equation \ref{eq:rza_error}), a practical quantity to estimate the validity of the approximation. Figure \ref{fig:rzaerrors_paper} shows $d^{\tm{RZA}}$ (blue) compared to the error on the initial positions that we would make without any Lagrangian reconstruction (red). In the latter case the error would be simply the displacement $|\bl \psi|$ itself.

The analysis reveals that a majority of the main haloes have a surprisingly low $d^{\tm{RZA}}$: the median is at 1.36 Mpc/$h$, the mean at 2.3 Mpc/$h$. This is well below the scale on which the first-order Zeldovich approximation is normally considered to be valid. Above all, it is a significant improvement over the ``0th order'' linear theory approximation that the overdensity peaks traced by the haloes do not move at all, leading to a mean error of $\langle|\bl \psi|\rangle = 8.7$ Mpc/$h$ for this particular simulation. The distribution of $d^{\tm{RZA}}$ is highly skewed: for most of the haloes, $d^{\tm{RZA}}$ is within a few Mpc/$h$, but at the same time a small fraction of objects has a very high $d^{\tm{RZA}}$. Because of this skewness, the median and the upper and lower quartiles are shown, being more meaningful than the mean and the $1\sigma$ interval. 

As expected, the success of RZA depends highly on the underlying overdensity. In higher-density regions, the non-linear enhancement of peculiar velocities is stronger, and shell-crossing occurs earlier, so that the Zeldovich approximation is not an optimal description of the dynamics there. Although we discarded the subhaloes, still some shell crossing will occur for the main haloes in the overdense regions. The dependence of $d^{\tm{RZA}}$ on the total velocity is also strong because high velocities are associated with dense environments. A couple hundred objects even have $d^{\tm{RZA}} > |\bl \psi|$, meaning that RZA completely fails there. They stick out in the bottom panel of Figure \ref{fig:rzaerrors_paper} as the three last bins with the highest density. All of those outliers are relatively low-mass objects  travelling at high velocities in the immediate vicinity of one of the most massive clusters in the box and thus experiencing significant non-linear contributions to their peculiar velocities. They can be removed by a scheme that is more rigorous than our shared-particles approach, such as increasing the minimum allowed distance to the next more massive haloes. In order to catch those extreme outliers, is thus sufficient to apply such a more rigorous scheme to the most dense environments only. In observational data, this  effectively means reducing rich galaxy clusters to a single data point, while keeping field galaxies ungrouped. If we would however apply such a more rigorous scheme to the whole data set, the retained objects would have a significantly lower $d^{\tm{RZA}}$ in average, but at the same time we would remove a substantial fraction of data points with useful information, which is undesirable.

We also found that there is only a very weak dependence of $d^{\tm{RZA}}$ on the actual mass of haloes. Furthermore, the displacement $|\bl \psi|$ shows no significant correlation with the mass either. This is consistent with the fact that the halo peculiar velocities $\bl u$ themselves are almost independent of halo mass \citep{Suhhonenko2003}. A simulation with larger boxsize would provide better statistics on the most massive objects, where a slight effect of smaller $\bl v$ can be seen, but they are not a focus of this work.

From this theoretical study on identified haloes in a cosmological simulation it is clear that the RZA can provide a reasonable estimate of the cosmological displacement and the initial position for most of the objects. The primary interest is now how well this scheme can be applied to more realistic observational data. This is discussed in the following section.

\section{Reconstruction from radial peculiar velocity data}
\label{sec:mocks}

In this section we study how well the RZA can be applied to a realistic observational data set. For this, we apply the RZA to mock data, drawn from the same BOX160 simulation as before, but featuring a limited data volume, observational errors, and knowledge of only the radial part $v_r$ of the peculiar velocity. In particular, we want to mimic two observational datasets: the Cosmicflows-1 catalogue \citep{Tully2009,Courtois2012}, which is available through the Extragalactic Distance Database\footnote{accessible online at edd.ifa.hawaii.edu} and features 1797 galaxy distances and peculiar velocities in 742 groups within 3000 km/s, and the upcoming Cosmicflows-2 catalogue \citep{Courtois2011a,Courtois2011b,Courtois2012arXiv,Tully2012arXiv}, which is expected to contain 7000 galaxy distances and peculiar velocities and extend out to 6000 km/s. This way we can directly estimate what quality of constrained simulations we can expect if we apply the RZA to these data sets, and by how much the reconstruction quality would improve with the upcoming new dataset.

\subsection{Building the mock catalogues}

In order to construct mock catalogues, we first have to choose a mock observer. We choose a galaxy group identified in BOX160, which consists of three main haloes with virial masses of 4.9, 6.0 and $6.7\times 10^{11} M_\odot/h$; we choose the middle one, calling it the simulated ``Milky Way''. The position of this halo, $\bl r_{\tm{MW}}$, marks the fixed position of the mock observer. The halo is at a distance of 17 Mpc/$h$ to the simulated BOX160 ``Virgo'', the next massive cluster. This distance is somewhat larger than the actual distance from the Milky Way to the centre of the Virgo Cluster (16 Mpc; see \citealt{Fouque2001}); however, this is not an issue here. To construct a mock catalogue, we cast a sphere with a fixed radius $R_{\tm{max}}$ around $\bl r^{\tm{MW}}$, and consider only haloes within this sphere -- the ``observational volume''. We choose a default value of $R_{\tm{max}} = 30$ Mpc/$h$, which mimics a redshift cut at 3000 km/s, similar to the Cosmicflows-1 catalogue. We also used a larger volume of $R_{\tm{max}} = 60$ Mpc/$h$, which resembles in quality the upcoming Cosmicflows-2 data. Since $\bl r^{\tm{MW}}$ is located near the centre of the 160 Mpc/$h$ simulation box, even at $R_{\tm{max}}=60$ Mpc/$h$ we are sufficiently far away from the box edge, so we will not suffer from the effects of the periodic boundary conditions. Our standard choice of the sphere within 30 Mpc/$h$ of the chosen MW candidate comprises 2.8\% of the total BOX160 volume and places the BOX160 ``Local Supercluster'' containing ``Virgo'' well inside the datazone. The larger 60 Mpc/$h$ sphere also encompasses the BOX160 ``Great Attractor'' and touches the simulated ``Coma'' and ``Perseus-Pisces'' clusters at its edge.

The $R_{\tm{max}} = 30$ Mpc/$h$ data volume leads to somewhat different conditions than considering the whole box, as we did in Section \ref{sec:rzaatz0}. In this data volume, there is a significant net displacement of about 4 Mpc/$h$. Also, the mock volume as a whole is overdense compared to the box average, and further non-linearily evolved than the average field at $z=0$. This is similar to the overdensity of the observed Local Universe region around the LSS, which is well established by observations \citep{Tully2008}. While the main haloes in BOX160 have a median displacement $|\bl \psi|$ of 8.7 Mpc/$h$ and a median RZA error $d^{\tm{RZA}}$ of 1.36 Mpc/$h$, for the 1243 main haloes inside the mock volume the median $|\bl \psi|$ is at 11.7 Mpc/$h$ and the median $d^{\tm{RZA}}$ at 2.8 Mpc/$h$. The net overdensity also means that there is a net inflow into the observational volume. It is interesting to see how the reconstruction will handle such a difficult case. In the Cosmicflows-1 data, the value of $H_0$ is chosen such that there is no net inflow/outflow with respect to the data zone \citep{Tully2008,Courtois2012}. This restriction may not be required in general. For the mock data, we keep the value of  $h=0.73$ as given by the simulation parameters.

Observationally, radial peculiar velocities $v_r^{\tm{pec}}$ and their errors are obtained from the measured galaxy distance $r$, some estimate of the absolute distance error $\Delta r$, and the observed redshift $v_r^{\tm{obs}}$ via
\begin{align}
v_r^{\tm{pec}} &= v_r^{\tm{obs}} - r \cdot H_0\;\;\;, \\
\Delta v_r^{\tm{pec}} &= - \Delta r \cdot H_0 \;\;\;\;\;\;\,\,.
\end{align}
To mimic these errors, we first take the known radial velocity of a halo from the AHF catalogue, $v_r^{\tm{AHF}}$, and the known distance $r^{\tm{AHF}}$ to the mock observer. In the Cosmicflows-1 catalogue, the different points come from different types of measurements with differently distributed errors between 7 and 20 \% \citep{Courtois2012}; here, we simplify the situation by assuming relative distance errors $\delta r$ that are Gaussian distributed with a constant $ (\delta r)_\tm{rms}$. If we choose a fixed value for this rms accuracy of the distance, then the absolute mock distance error is generated via
\begin{align}
\Delta r^{\tm{mock}} = \tm{G}(0,1) \cdot (\delta r)_\tm{rms} \cdot r^{\tm{AHF}}\;\;\;,
\end{align}
where G(0,1) is a random number drawn from a Gaussian distribution with mean 0 and variance 1. Then, the radial velocity with the added error for the mock catalogue is computed as follows:
\begin{align}
v_r^{\tm{mock}} &=  v_r^{\tm{AHF}} + \Delta v_r^{\tm{mock}} \;\;\;,\\
\Delta v_r^{\tm{mock}} &= - \Delta r^{\tm{mock}} \cdot H_0 \;\;\;\;\;\,\,.
\end{align}
This data forms the WF/CR input constraints, where $i = 1, \, \ldots \, N$ goes over all haloes in the mock catalogue.

In observational data, $v_r^{\tm{obs}}$ is observed in the rest frame of the observer and is then transformed to the rest frame of the CMB dipole. This is preferred cosmological frame of reference within which the primordial cosmological perturbations are analyzed and cosmological simulations are conducted.

For the mocks, we achieve a similar setup by not considering the peculiar motion of the simulated MW halo where we placed the mock observer. We rather take directly the velocities in the fixed rest frame of the simulation box as computed by AHF.

Our ``standard'' mock catalogue is the C30\_10, which we consider a ``typical'' sparse peculiar velocity dataset. We take the procedure of considering only main haloes as a proxy for the ``grouping'' performed on observational data. The C30\_10 contains all main haloes above a mass cut $M_\tm{min} = 10^{11.9} M_\odot/h$ within $R_{\tm{max}} = 30$ Mpc/$h$, yielding 588 radial velocity datapoints. This choice gives the C30\_10 similar properties to the grouped Cosmicflows-1 catalogue, but somewhat more sparse\footnote{Choosing this sample was motivated by the fact that at the time we commenced this study, we had a preliminary version of the Cosmicflows-1 available that contained exactly the same number of 588 galaxy groups within 30 Mpc/$h$. The current version of the Cosmicflows-1 catalogue is less sparse with 742 galaxy groups within the same volume.}. For the ``observational'' rms distance error $(\delta r)_\tm{rms}$ of this mock catalogue we choose a value of 10\%. This is similar to the observational data: while the median rms distance error is somewhat higher at 13\% on the individual galaxies in Cosmicflows-1, this error reduces when the galaxies are arranged in groups. The C30\_10 is interesting because even if the data are improving in terms of the number of individual galaxy distances, the number of galaxy groups in a radius of 30 Mpc/$h$ is probably not going to vary by much, nor is the accuracy on the most nearby distances. Note that a rms distance error of $(\delta r)_\tm{rms} = 10\%$ leads to a relatively high error on the peculiar velocities: at a distance of $r = 30$ Mpc/$h$, this corresponds to an rms error of 300 km/s on the peculiar velocity, which is approximately equal to the variance $\sigma$ of the peculiar velocity values.

In order to also have a less sparse sample, we also use a lower mass cut at $M_\tm{min} = 10^{11.5} M_\odot/h$; the corresponding mock is labelled E30\_10 and contains 1243 radial velocities within $R_\tm{max} = 30$ Mpc/$h$. We also use a mock with the larger volume of $R_{\tm{max}} = 60$ Mpc/$h$, labelled E60\_10, mimicking the upcoming Cosmicflows-2 data. This mock features 7637 peculiar velocities. Both of these mocks also feature the same kind of distance errors with $(\delta r)_\tm{rms} = 10\%$.

\subsection{Application of the RZA to mock data}

In order to apply RZA reconstruction of the displacement field from the mock catalogues, we first need to estimate the three-dimensional peculiar velocity vector $\bl v$ from its given radial component $v_r$ while simultaneously accounting for the uncertainties of the $v_r$ values due to the mock observational errors. For this, we use the Wiener Filter (WF) reconstruction method introduced in \ref{subsec:rza_wf}. The radial peculiar velocities in the mock catalogues are very well approximated by a Gaussian distribution, therefore the WF is a very good estimator \citep{Zaroubi1999}. As the prior model we use the same WMAP3 power spectrum that was used to set up the BOX160 simulation. It is one of the strengths of the WF method that it can not only filter out noise and extrapolate/interpolate the field into unsampled regions from sparse and noisy datasets, but it can also recover any linear functional of the overdensity $\delta$ (here, the three-dimensional velocity field $\bl v$) from a sampling of any other linear functional (here, the radial velocity component $v_r$).

Using the radial velocity data points from the mock catalogue as constraints $c_i$, we evaluate Eq. \ref{eq:vwf} in order to reconstruct all three cartesian components of $\bl v$. Note that we have to evaluate Eq. (\ref{eq:vwf}) only at the positions $\bl r_i$ of the mock data points, which significantly reduces the computational cost compared to a reconstruction of the entire field. We can then continue and apply the RZA. The only difference from the approach in Section \ref{sec:rzaatz0} is that, since the true values of $\bl v$ are unknown, we now take the WF estimate $\bl v^\tm{WF}$ instead. This technique is straightforward to apply also to a real observational dataset.

We perform the RZA reconstruction on the mock datasets by evaluating Eqs. (\ref{eq:psi_rza}) and (\ref{eq:xinit_rza}) to recover $\bl \psi^\tm{RZA}$ and $\bl x_\tm{init}^\tm{RZA}$ for each object in the mock dataset, only this time using the $\bl v^\tm{WF}$ estimate for $\bl v$. Since we have the initial conditions of the simulation available, we can check the results and also evaluate the RZA error $d^\tm{RZA}$ via Eq. (\ref{eq:rza_error}) for the different mock catalogues to estimate the accuracy of the reconstruction. 

\subsection{Reconstructed displacements}

\begin{figure}
\centering
\includegraphics[scale=0.65]{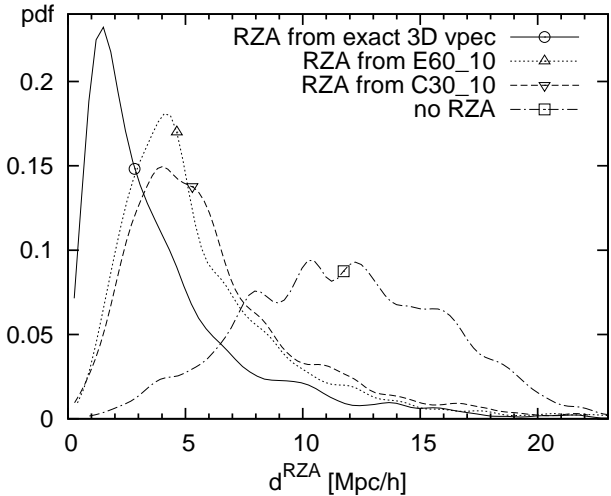}
\caption{Distribution function of the RZA displacement error $d^{\tm{RZA}}$ for main haloes within 30 Mpc/$h$ from the BOX160 mock observer for RZA reconstructions from the exact 3D halo peculiar velocities (solid), the two radial peculiar velocity mocks E60\_10 (dotted) and C30\_10 (dashed), and no reconstruction at all, i.e.\ $d = |\bl \psi|$ (dot-dashed). The symbols are placed at the median of each distribution. Each pdf was convolved with a 0.5 Mpc/$h$ Gaussian kernel to obtain a smooth curve.}
\label{fig:kernelpdf_paper}
\end{figure}

Figure \ref{fig:kernelpdf_paper} shows the RZA displacement error $d^{\tm{RZA}}$ distribution of the RZA reconstruction obtained for the two mocks E60\_10 and C30\_10  with the WF technique described above. Both mocks feature only the radial component of the halo peculiar velocity and observational errors of $\delta r = 10\%$. The RZA displacement error $d^{\tm{RZA}}$ resulting from this procedure is compared to the  $d^{\tm{RZA}}$ in the case if we know the full three-dimensional peculiar velocity vectors of the haloes with no error, like in the analysis in Section \ref{sec:rzaatz0}, and to the case of no Lagrangian reconstruction (then, the error is simply $|\bl \psi|$, the amplitude of the displacement itself). The $d^{\tm{RZA}}$ values considered for Figure \ref{fig:kernelpdf_paper} are only those of the haloes inside the mock data volume, i.e.\ a sphere of 30 Mpc/$h$ radius around the mock observer at $\bl r_\tm{MW}$. This volume is an overdense region compared to the whole computational box of the BOX160 simulation. Therefore, the mean displacements and peculiar velocities are higher. In this mock volume, the median displacement field is 11.7 Mpc/$h$, which is also the median error of the estimated initial positions if one does not use Lagrangian reconstruction. With perfect knowledge of the 3D halo peculiar velocities $\bl v$, RZA reconstruction reduces this error to $d^{\tm{RZA}}=2.8$ Mpc/$h$. For the mocks, which contain only the radial component $v_r$ of $\bl v$ and are sparse and noisy, the WF + RZA method produces a median $d^{\tm{RZA}}$ of typically around 5 Mpc/$h$ (5.30 Mpc/$h$ for the C30\_10 mock and 4.64 Mpc/$h$ for the better-quality E60\_10 mock). This is still a significant improvement over the previous method not using Lagrangian reconstruction. The increased datapoint density and data volume in the E60\_10 mock leads to only a small improvement in reconstruction quality within 30 Mpc/$h$ of $\bl r_\tm{MW}$. This confirms that the Wiener filter is capable of producing an already good estimate of the true underlying field from very sparse, noisy, and incomplete data.

\begin{figure}
\centering
\includegraphics[width=7.5cm]{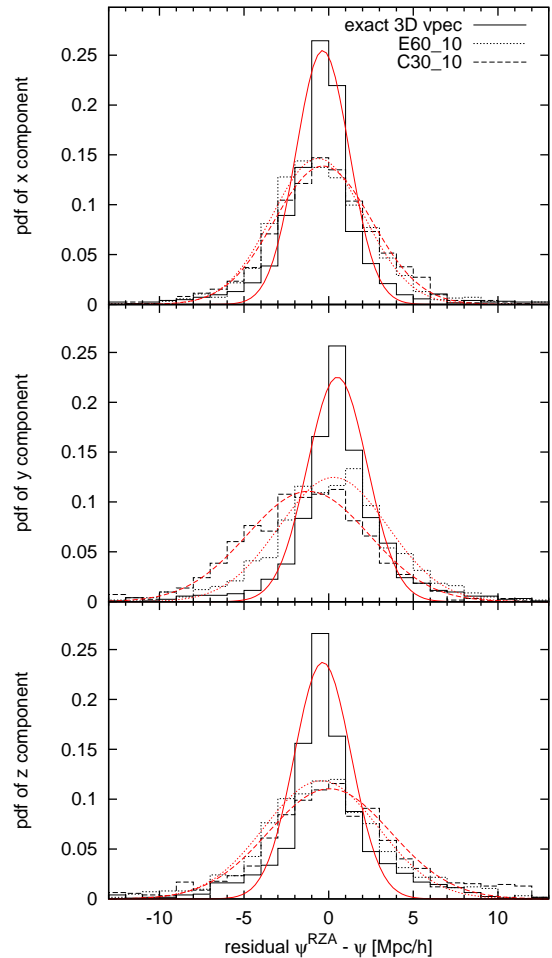}
\caption{
The distribution of the three Cartesian components of the residual  $\bf{\psi}^{\tm{RZA}}-\bf{\psi}$ for the main haloes within 30 Mpc/$h$ from the BOX160 mock observer for RZA reconstructions from the exact 3D halo peculiar velocities (solid) and the two radial peculiar velocity mocks E60\_10 (dotted) and C30\_10 (dashed). The red curves show the normal distributions fitted for the actual distributions.}
\label{fig:drza-3d}
\end{figure}

Figure \ref{fig:drza-3d} shows the distribution of the error in the RZA displacement for the three Cartesian components for three  mock catalogs. The catalogs consist of the case of exact 3D halo peculiar velocities   and the two radial peculiar velocity mocks, E60\_10  and C30\_10. For all cases and for the three different components the distribution of the residual is well fitted by a normal distribution. The anisotropic distribution of the data points of the mock catalogs leads to an anisotropic distribution of the residuals. For the particular constrained BOX160 simulation, from which the mock catalogs are drawn, the distributions of the residuals in the X and Z directions are very similar, and both are distinct from the distribution in the Y direction.

\begin{figure*}
\centering
\includegraphics[trim=0cm 0cm 0cm 0cm, clip=true, scale=0.62]{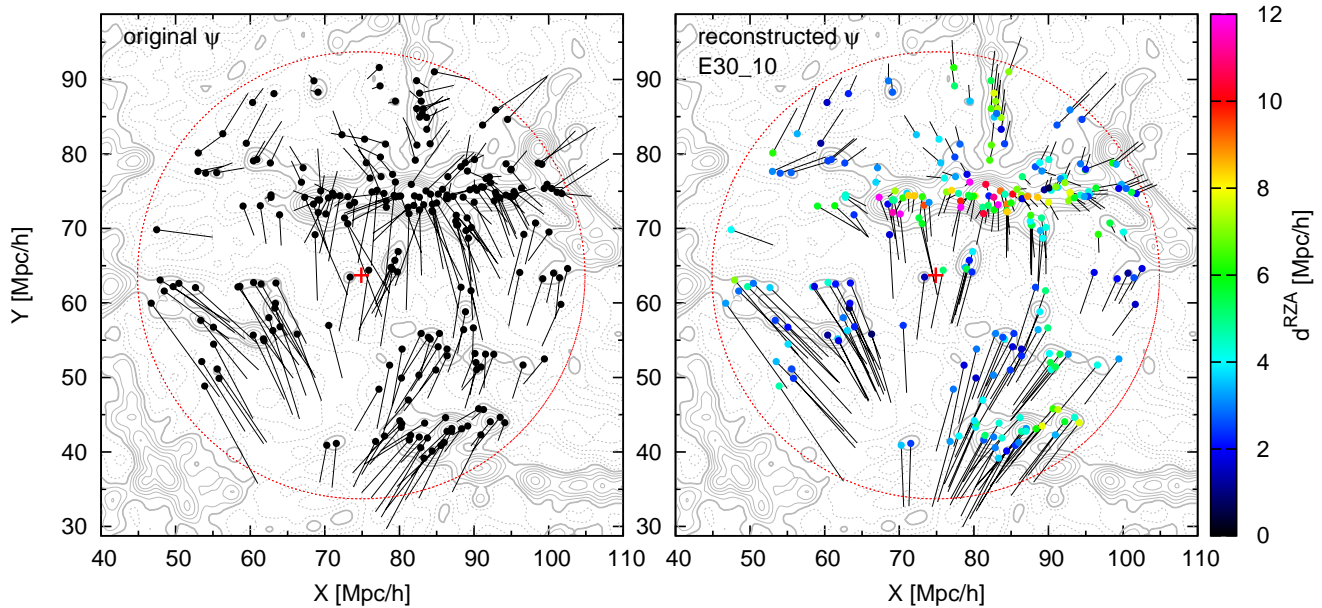}
\caption{Actual BOX160 displacement field (left) vs. RZA reconstruction from radial peculiar velocities (using mock E30\_10, with mock observational distance errors of 10\%). The dots show the $z=0$ position $\bl r$ of the haloes contained in the mock data; the lines show the displacement vector $\bl \psi$ from the initial conditions. In the reconstruction, the RZA error $d^{\tm{RZA}}$ is highlighted by colour-coding of the points. The contours show the underlying dark matter density (smoothed with a 1.5 Mpc/$h$ Gaussian). Shown is a 15 Mpc/$h$ thick slice containing the simulated BOX160 Local Supercluster with Virgo at X=82, Y=74. The red cross marks the position of the mock observer.}
\label{fig:mockrzamap}
\end{figure*}

Figure \ref{fig:mockrzamap} shows the actual displacement field $\bl \psi$ and the reconstructed $\bl \psi^\tm{RZA}$ for the E30\_10 mock in a 15 Mpc/$h$ thick slice through the mock volume. The comparison reveals that although the input data consisted only of the radial components of the haloes' peculiar velocity vectors, the Wiener filter + RZA method manages to recover a good estimate (right panel of Figure \ref{fig:mockrzamap}) of the true three-dimensional displacement field (left panel of Figure \ref{fig:mockrzamap}). The RZA error $d^{\tm{RZA}}$ values for the data points of this mock are color-coded in the right panel. The reconstruction is very accurate ($d^{\tm{RZA}} \lesssim 2$ Mpc/$h$) in the less dense regions of the mock outside the very dense ``supercluster'' structure above the centre. Inside the supercluster however, the original displacements are generally not recovered by this procedure. This can be explained because, even after the grouping procedure, peculiar velocities inside such dense environments are dominated by non-linear effects. In order to prevent the occurence of high $d^{\tm{RZA}}$ values above 5 Mpc/$h$, one would have to enforce a much stronger grouping inside such dense clusters.

\section{Constraining initial conditions with RZA}
\label{sec:ics}

In the last section we established that using a realistic peculiar velocity dataset, the RZA can give good estimates of the underlying cosmological displacement field and the initial position of the datapoints. The main goal remains, however, to use such data to constrain cosmological initial conditions and to run constrained simulations. We therefore have to fit the RZA into the CR method (Section \ref{subsec:rza_cr}) that we use to generate constrained ICs. The scheme has to be designed around the basic requirement that the input for the CR algorithm that produces Gaussian ICs has to consist of a set of constraints $\Gamma = \{ c_i\}$. 

\subsection{Method}

The RZA allows us to improve upon the previously used method, which consisted of taking the radial peculiar velocity values $v_{r,i}$ at their observed positions $\bl r_i$ at $z=0$ and using them as constraints $c_i$ for the ICs at $z_\tm{init}$. (In the following, we refer to this as ``Method I''.) One could think of now taking as constraints the reconstructed displacement fields $\bl \psi^\tm{RZA}$ at the reconstructed initial positions $\bl x^\tm{RZA}_\tm{init}$ instead. Essentially the same approach was employed by \citet{Lavaux2010b} to construct constrained initial conditions, except that the reconstructed displacement fields and initial positions resulted from a MAK reconstruction procedure applied to the 2MASS redshift survey. However, in our case this approach will not work. The displacement field $\bl \psi^\tm{RZA}$ was obtained with the WF/CR algorithm, and using its results as constraints for a CR would mean to apply the WF/CR algorithm a second time, i.e.\ \emph{iteratively} to the same data set. This results in an unacceptable loss of power due to the conservative nature of the WF, and the additional information gained by performing a Lagrangian reconstruction would be smoothed away by the filter. It is a general rule that applying the WF iteratively does not lead to meaningful results.

To overcome this obstacle, we tried different modifications of this idea. The optimal strategy that we found consists of displacing the radial velocity datapoints $v_r$ from the (mock) catalogue from their observed position $\bl r$ at $z=0$ ``back in time'' to the RZA-reconstructed initial position $\bl x_\tm{init}^\tm{RZA}$, while leaving all other properties attached to the datapoints unchanged (the direction of the constrained component of $\bl v$, its amplitude, and the associated observational error). In the following, we refer to this as ``Method II''. (A ready-to-use implementation of this shifting procedure is included in the \textsc{ICeCoRe} code.) We then use this shifted set of constraints as input to generate a new CR. In this case, we are not applying the WF/CR iteratively, since we use the first (WF) run only to shift the datapoint positions, before we use them for the second (CR) run. We therefore change only the positions $\bl r_i$ of the original constraints $c_i$ to obtain the new constraints $c_i^\tm{RZA}$ that are actually used for generating ICs.

Note that for the new constraints $c_i^\tm{RZA}$, the constrained component is no longer the radial component with respect to the observer position $\bl r_\tm{MW}$, because the position of the constraint was shifted from its observed position $\bl r$ to its estimated initial position $\bl x_\tm{init}^\tm{RZA}$. For this, we used our new \textsc{ICeCoRe} code\footnote{Our previously used numerical implementation of the CR algorithm was restricted to constraining the component that was the radial direction at its position (i.e. parallel to the position vector w.r.t. the observer). \textsc{ICeCoRe} removes this restriction and allows a much more flexible placement of different types of constraints.}. 
It allows to constrain arbitrarily directed components $v_\mu$ (along any unit vector $\hat{\bl e}_\mu$) of the peculiar velocity vector $\bl v$  for applying the WF/CR operator in Eq. (\ref{eq:deltacr}), which is straightforward to implement using Eqs. (\ref{eq:mucomp}) and (\ref{eq:munucomp}). This however adds the requirement to explicitly specify for each velocity constraint the unit vector $\hat{\bl e}_\mu$ of the component to be constrained. In our case, the unit vectors $\hat{\bl e}_\mu$ of the constrained peculiar velocity components $v_\mu$ will correspond to $\hat{\bl e}_r$ for the observable radial direction $v_r$ with respect to the observer at $z=0$, however the positions $\bl x$ of the constraints are not those at $z=0$, but displaced to the RZA-estimated initial position $\bl x_\tm{init}^\tm{RZA}$ according to Eq. (\ref{eq:xinit_rza}), so that  $\hat{\bl e}_r$ is not parallel to $\bl x_\tm{init}^\tm{RZA}$.

In practice, the steps for obtaining constrained ICs for $N$-body simulations using the RZA method are as follows:
\begin{enumerate}
\item
Apply a grouping to the input radial peculiar velocity data to ``linearize'' it and to filter out virial motions. The resulting set of radial peculiar velocities $v_{r,i}$ at positions $\bl r_i$ defines the set of constraints $c_i$.
\item
Run a WF reconstruction on the $c_i$ by evaluating Eq. (\ref{eq:deltawf}) at the positions $\bl r_i$ This will yield an estimate of the three-dimensional velocities $\bl v_i$ at $\bl r_i$.
\item
Apply the RZA by evaluating \ref{eq:xinit_rza} to yield an estimate of the initial positions $\bl x_\tm{init}^\tm{RZA}$.
\item
Build a new set of constraints $c_i^\tm{RZA}$ by shifting the positions of the $c_i$ back in time from $\bl r_i$ to $\bl x_\tm{init}^\tm{RZA}$.
\item
Run the CR algorithm, using the constraints $c_i^\tm{RZA}$, to obtain a realization of constrained initial conditions, scale it to the desired redshift $z_\tm{init}$, and run a constrained $N$-body simulation.
\end{enumerate}
The implementation details of how to set up particle initial conditions from the $\delta^\tm{CR}$ field are not different from conventional cosmological simulations, see e.g. \citet{Prunet2008,Gottloeber2010arXiv}.

With this scheme, it is possible to generate significantly more accurate constrained simulations from radial peculiar velocity catalogues than with the previous method without Lagrangian reconstruction. In paper III of this series we will investigate such constrained simulations in detail.

\section{Summary, conclusions and outlook}
\label{sec:summary}

This work is the first in a series of papers investigating the problem of generating initial conditions for constrained simulations of the Local Universe using galaxy peculiar velocities. Our main motivation is to extend the previously used Constrained Realizations (CR) method to include a Lagrangian reconstruction scheme in order to improve the quality of the resulting constrained simulations. In this paper, we propose a method to reconstruct the cosmological displacement field from observable peculiar velocities. We find that for this task, the relatively simple Zeldovich approximation (ZA) is a very powerful approach when applied directly to peculiar velocities, instead of applying it to a sampling of the density field as it is commonly done in various other contexts. Instead of using the ZA to approximate the cosmic structure formation forward in time, which was its original motivation, we use it in the reverse time direction to estimate the cosmological initial conditions at high $z$ in the linear regime from the peculiar velocities at $z=0$. We call this approach the Reverse Zeldovich Approximation (RZA).

As the main finding of this work we conclude that, for the task of generating constrained realisations of the Local Universe from peculiar velocity data, a Lagrangian reconstruction scheme such as the RZA reconstruction presented here provides a significant improvement over treating the peculiar velocities with linear theory, which was the previous approach. We tested this method with realistic radial peculiar velocity mock catalogues drawn from a cosmological reference simulation (which itself is a constrained simulation of the Local Universe). For a reconstruction of the three-dimensional galaxy peculiar velocities from the radial components in the mock catalogues, we used the well-established Wiener filter method. We confirm that it performs very well in compensating for the radial component limitation and the observational errors. We found that directly applying the WF/CR operator to peculiar velocity data at $z=0$ is actually not a very good estimate of the initial conditions due to the displacements, the non-linearities of the data, and the relatively large systematic errors. However, the Wiener filter reconstructed 3D peculiar velocity field is a reasonably accurate estimate of the displacement field $\bl \psi$, with a typical error of a few Mpc/$h$. This allows us to generate a better estimate of the initial positions of the data points' progenitors. We find that for halo peculiar velocities, the RZA is able to recover the correct initial positions with a median error of only 1.36 Mpc/$h$, where we use the BOX160 simulation as a reference universe. For the realistic mock catalogues drawn from this simulation (featuring various observational errors) this median increases to 
$5 \ \hMpc$. This is a significant improvement over the previous approach of neglecting the displacement field, which introduces errors on a scale of 
$10\  \hMpc$ or even higher. 

In paper II of this series, we will investigate in detail how much the individual observational errors and limitations affect the result of the RZA reconstruction, in order to derive what observational data sets would be ideally suited for a reconstruction of the ICs of our Local Universe. We can already make a first prediction based on the work presented here. Distance catalogues containing spiral galaxies, such as Tully-Fisher data, may be a better choice than data covering early-type galaxies such as the fundamental plane and surface brightness fluctuations methods. Spiral galaxies provide a more homogeneous mapping of the sky and are less biased towards high-density regions where non-linear effects become stronger. These regions are exactly where the RZA reconstruction method fails (the displacement error becomes very high due to non-linearities). It will also be important to optimize the data grouping method for filtering out virial motions. Our results suggest that the best strategy is to reduce each group of data points that form a virialised or otherwise strongly gravitationally interacting structure into a single data point, in order to remove non-linear virial motions completely. Testing the performance of different input data in this context will be the subject of further studies.

The displacement and initial position estimate generated by the RZA reconstruction can be used to constrain initial conditions (ICs) for constrained cosmological simulations. These constrained ICs can be set up in a second WF/CR operator step, using a new set of constraints that is obtained by shifting the original datapoints to their RZA-estimated initial positions. Constrained simulations can be run by feeding the result of that procedure as the ICs for an $N$-body simulation code. In paper III of this series, we will analyze in detail the accuracy and limitations of such constrained simulations and quantify the increase in reconstruction quality that can be obtained by using RZA.

We are also exploring the extension of the RZA reconstruction to higher-order Lagrangian perturbation theory. However, any method of higher order than the first-order Zeldovich approximation will break the locality of the approximation, and the whole field has to be considered instead of the given discrete data points. This will invariably introduce additional systematic errors. It is a very difficult task to overcome this obstacle if one is presented with very sparse and inhomogeneously sampled input data. With the first-order RZA, we therefore may already have found a near-optimal method of reconstructing initial conditions from observational galaxy radial peculiar velocity data.


\section*{Acknowledgments}

TD would like to thank R. Brent Tully, Matthias Steinmetz, Francisco-Shu Kitaura, Jochen Klar, Adrian Partl, Steffen Knollmann, Noam I Libeskind, Steffen Hess, Guilhem Lavaux, Saleem Zaroubi, and Alexander Knebe for helpful and stimulating discussions.
YH and SG acknowledge support by DFG  under GO 563/21-1.
YH has been partially supported by the Israel Science Foundation (13/08).
TD and SG acknowledge support by DAAD for the collaboration with H.M. Courtois and R.B. Tully.
We would like to thank the referee of this series of papers for her/his very careful and fast reading of the manuscripts and the many constructive comments which improved the three papers substantially.

\bibliography{Doumler2012_paper1.bib} \bsp

\label{lastpage}

\newpage

\end{document}